%% file: Paper-1607.tex
\documentclass[runningheads]{llncs}
\usepackage[T1]{fontenc}
\usepackage{graphicx,verbatim}
\usepackage{cite}
\usepackage{amssymb}
\usepackage{bm}
\usepackage{multirow, booktabs}
\usepackage{amsmath}
\usepackage{utfsym}
\usepackage[pagebackref,breaklinks,colorlinks,allcolors=blue]{hyperref}
\usepackage{subfig}
\usepackage[misc]{ifsym}
\usepackage{color}

\begin{document}

\title{Leveraging Semantic Asymmetry for \\ Accurate Gross Tumor Volume Segmentation of Nasopharyngeal Carcinoma in Planning CT}

\author{Zi Li\thanks{Equal contribution. \textrm{\Letter} Corresponding author.}\inst{1,2,4}$^{(\textrm{\Letter})}$ % index{Li, Zi}
\and Ying Chen$^{\star}$\inst{3} % index{Chen, Ying}
\and Zeli Chen\inst{1,4} % index{Chen, Zeli}
\and Yanzhou Su\inst{1,4} % index{Su, Yanzhou}
\and Tai Ma\inst{1} \and \\ % index{Ma, Tai}  
Tony C. W. Mok\inst{1,4} % index{Mok, Tony}
\and Yan-Jie Zhou\inst{1,4} % index{Zhou, Yan-Jie}
\and Yunhao Bai\inst{1} % index{Bai, Yunhao}
\and Zhinlin Zheng\inst{1,4} % index{Zheng, Zhinlin}
\and Le Lu\inst{1} \and % index{Lu, Le}
\\ Yirui Wang\inst{1} % index{Wang, Yirui}
\and Jia Ge\inst{3} % index{Ge, Jia}
\and Senxiang Yan\inst{3} % index{Yan, Senxiang}
\and Xianghua Ye\inst{3}$^{(\textrm{\Letter})}$ % index{Ye, Xianghua}
\and Dakai Jin\inst{1} % index{Jin, Dakai}
}
\authorrunning{Z. Li et al.}
\titlerunning{Leveraging Semantic Asymmetry for NPC in Planning CT}

\institute{DAMO Academy, Alibaba Group \and
The University of Hong Kong,  Hong Kong \and
The First Affiliated Hospital, Zhejiang University, China  \and Hupan Lab, 310023, Hangzhou, China \\
\email{alisonbrielee@gmail.com;hye1982@zju.edu.cn}
}

\maketitle              % typeset the header of the contribution
\begin{abstract}
In the radiation therapy of nasopharyngeal carcinoma (NPC), clinicians typically delineate the gross tumor volume (GTV) using non-contrast planning computed tomography to ensure accurate radiation dose delivery. However, the low contrast between tumors and adjacent normal tissues requires radiation oncologists to delineate the tumors with additional reference from MRI images manually. 
In this study, we propose a novel approach to directly segment NPC gross tumors on non-contrast planning CT images, circumventing potential registration errors when aligning MRI or MRI-derived tumor masks to planning CT. To address the low contrast issues between tumors and adjacent normal structures in planning CT, we introduce a 3D Semantic Asymmetry Tumor Segmentation (SATS) method.
Specifically, we posit that a healthy nasopharyngeal region is characteristically bilaterally symmetric, whereas the presence of nasopharyngeal carcinoma disrupts this symmetry. 
Then, we propose a Siamese contrastive learning segmentation framework that minimizes the voxel-wise distance between original and flipped areas without tumor and encourages a larger distance between original and flipped areas with tumor.
Thus, our approach enhances the sensitivity of deep features to semantic asymmetries.
Extensive experiments demonstrate that the proposed SATS achieves the leading NPC GTV segmentation performance in both internal and external testing.
\end{abstract}

\input{sec/1_intro}

\input{sec/2_method}

\input{sec/3_experiment}

\section{Conclusion}
We propose a novel semantic asymmetry learning method that leverages the inherent asymmetrical properties of tumors in the nasopharyngeal region, thereby enhancing the accuracy of nasopharyngeal carcinoma tumor segmentation. 
Our method demonstrates a significant improvement in NPC GTV segmentation by effectively utilizing semantic symmetry inherent in anatomical structures, achieving superior performance compared to state-of-the-art methods, as validated on both an internal test set and an independent external dataset.

\bibliographystyle{splncs04}
\bibliography{mybib}

\end{document}

%% file: sec/1_intro.tex
\section{Introduction}

Nasopharyngeal carcinoma (NPC) ranks among the most prevalent head \& neck malignancies affecting the nasopharyngeal region, with patient prognosis substantially enhanced through early diagnosis and intervention~\cite{chua2016nasopharyngeal}. A significant proportion of NPC patients can achieve complete remission following radiation therapy (RT)~\cite{chen2019nasopharyngeal}. 
Notably, this type of cancer exhibits a remarkable sensitivity to radiation therapy, wherein a pivotal component of this therapeutic intervention is the accurate delineation of the gross tumor volume~(GTV). 
In clinical practice, magnetic resonance imaging (MRI) has emerged as the predominant imaging modality for NPC, owing to its superior resolution in visualizing soft tissues. Subsequently, cross-modality registration is conducted between MRI and non-contrast planning computed tomography (pCT) to transfer tumor delineations from MRI to pCT scans for treatment planning~\cite{razek2012mri}. However, cross-modality registration is non-trivial due to substantial modality gaps and variations in scanning ranges. 
Alternatively, physicians may integrate pCT and MRI mentally to assist in delineating the GTV. Nevertheless, this approach is time-consuming, often taking 1-2 hours per case, and is prone to potential inaccuracies.

Recent advances in learning-based methods have shown success in segmenting NPC tumors from MRI scans~\cite{huang2019achieving,ke2020development,luo2021efficient,liao2022automatic,9684475,luo2023deep}. However, MRI does not provide direct electron density measurements, which are critical for radiotherapy planning. Tumor masks derived from MRI must be transformed into pCT using image registration, a process prone to alignment errors. 
Approaches~\cite{ma2019nasopharyngeal,ChenQYLLLGW20} combine CT and MRI for tumor segmentation, although misalignment between the two modalities can reduce performance compared to single-modality approaches.
Other researchers~\cite{men2017deep,li2019tumor,mei2021automatic,bai2021deep} focus on contrast-enhanced CT-based segmentation. Still, these methods often achieve low performance (e.g., Dice scores below 70\%) due to tumor infiltration into adjacent tissues and limited contrast in pCT, particularly for soft tissues such as mucous membranes, muscles, and nerves.

This study aims to segment the NPC gross tumor volume (GTV) in non-contrast planning CT (pCT), avoiding registration errors associated with aligning MRI or MRI-derived tumor masks to pCT. Direct segmentation of NPC GTV in non-contrast pCT is challenging due to indistinct boundaries between tumors and adjacent soft tissues~\cite{li2024improved}, such as membranes, muscles, and vessels.
Meanwhile, we observe that medical image analysis benefits from the bilateral symmetry of human anatomy, evident in structures like the head, brain, breasts, lungs, and pelvis. Research~\cite{liu2019using,LiuZZLZZLWY19,ChenWZLCHXHLLM20,HuangLLW22,zeng2023two,huang2024lidia} highlights the utility of symmetry-based approaches in enhancing early detection capabilities.
Therefore, we propose a tumor segmentation method for NPC, which leverages the observation that a healthy nasopharyngeal region is bilaterally symmetric, but the presence of a tumor disrupts this symmetry.  
While prior work has explored symmetry in medical imaging, our approach differs significantly in how symmetric features are utilized. For instance, \cite{HuangLLW22} employs symmetric position encoding for brain structures using an autoencoder, without explicit constraints on symmetric or asymmetric regions (e.g., via custom losses or modules). \cite{zeng2023two} leverages pelvic symmetry to detect fractures but focuses primarily on symmetric anatomy. In contrast, our method emphasizes the contrast between asymmetric lesion areas.

The main contributions of this work are:
1) We introduce a 3D \emph{semantic asymmetry tumor segmentation (SATS)} method for NPC GTV in non-contrast pCT, which is the most common imaging modality in RT planning. To the best of our knowledge, this is the first work to tackle the NPC GTV segmentation in non-contrast CT scans and employ the symmetry cue for the GTV segmentation.
2) We develop a Siamese contrastive learning segmentation framework with an asymmetrical region selection approach, which facilitates the learning of asymmetric tumor features effectively.
3) We demonstrate that our proposed SATS achieves \emph{state-of-the-art} performance in NPC GTV segmentation, outperforming the leading methods in internal and external testing datasets.

%% file: sec/2_method.tex
\section{Method}
We propose a 3D semantic asymmetry tumor segmentation (SATS) method based on the semantic asymmetry property of the gross tumor in the nasopharyngeal area, to enable accurate NPC GTV segmentation. 
Given one CT scan, as shown in Figure~\ref{fig:overview}~(a), we utilize a shared encoder-decoder module to process both the original image $I \in \mathbb{R}^{D\times H\times W}$, where $D, H, W$ are CT image spatial dimensions, and its flipped image $I'$, thereby encoding them into a symmetric representation. 
Subsequently, we introduce a non-linear projection module and a distance metric learning strategy to refine the resulting feature maps. We intend to maximize the dissimilarity between $E$ and $E_{f}$ at corresponding anatomical locations on the abnormalities and normalities. The distance metric learning paradigm is illustrated in Figure~\ref{fig:overview}~(b).

\subsection{Asymmetrical Abnormal Region Selection}
\begin{figure*}[!t]
\begin{center}
\includegraphics[width=1\linewidth]{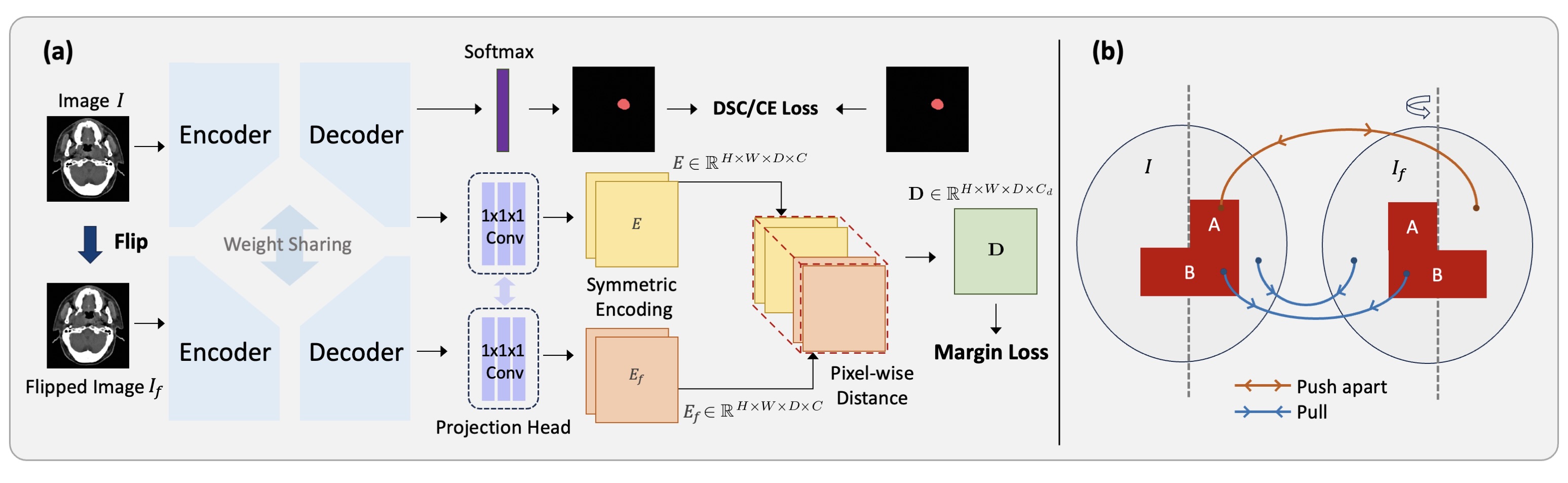}
\end{center}
\caption{\textbf{(a)} Our SATS model begins with the encoder-decoder module, which uses shared weights to process two input signals and encode them into a discriminative representation. This representation is then further processed through a non-linear projection module and a distance metric learning module to produce feature maps.
\textbf{(b)} A graphical representation of our metric learning strategy.
Circles indicate individual CT images, $I$, while red squares highlight the tumors. The tumors are composed of A and B, representing symmetrical and asymmetrical lesions relative to the central sagittal axis of symmetry, respectively. }
\label{fig:overview}
\end{figure*}

We focus on asymmetrical lesion areas relative to the central sagittal axis, i.e., region B of Figure~\ref{fig:overview}~(b). To this end, we perform: 1) head-neck position normalization (bilateral symmetry along the central sagittal axis) of the overall head-neck region by utilizing rigid registration (rotation and translation). 2) The asymmetrical abnormal region is obtained by subtracting symmetrical lesion regions from the original mask.

To be specific, considering that image asymmetry may originate from pathological or non-pathological sources, such as changes in imaging angles and patient postures, we pre-process the CT scans using \cite{Tian2023SAMEAS} to ensure that the scans are symmetric along the central sagittal axis. We manually select a patient CT image with bilateral symmetry along the central sagittal plane, which serves as an atlas, and then align other patient CT images to the atlas space through affine registration. This step helps to alleviate the influence of other asymmetric anatomical structures in the head \& neck that may mislead the model.

Then, we detect asymmetric abnormal regions using the available tumor annotation. 
The semantic segmentation mask of $I$ is denoted as $s\in\{0,1\}^{D\times H \times W}$, where $0$ represents the background and $1$ represents the foreground of tumors. Through the flip operation, we can obtain the flipped semantic mask $s'$ of $I'$.
Subsequently, an asymmetrical mask $\bm{m}$ is defined to locate asymmetrical regions in the image $I$, as $\bm{m}= \bm{s} - \bm{s} \cap \bm{s'}$, where $\bm{m}\in\{0,1\}^{D\times H \times W}$. Note that $1$ and $0$ represent the asymmetrical and symmetrical regions in $I$, respectively. 

\subsection{Asymmetrical Learning Strategy}
Our segmentation loss function is comprised of two components: a combination of Dice and entropy loss for the conventional segmentation purpose, and a voxel-wise margin loss specifically designed for asymmetric abnormal regions.

\textbf{Metric-based margin loss.} 
In the asymmetric anomaly region, we aim to minimize the similarity between the features of any point and its corresponding point on the central sagittal axis. To achieve this, we employ pixel-level margin loss. Based on above asymmetrical abnormal region $\bm{m}$, the margin loss between features $E \in \mathbb{R}^{H\times W\times D\times C}$, where $C$ is the number of output features, and flipped $E'$ after a non-linear projection is as:
\begin{equation}
\begin{aligned}
    \bm{l}_{margin} = {\textstyle \sum_{i,j,k}^{D,H,W}} [ \mathbf{1}_{(m(i,j,k)=1)}|| E(i,j,k) -E'(i,j,k) ||^2 + \\ \mathbf{1}_{(m(i,j,k)\neq 1)} \max (t - || E(i,j,k) -E'(i,j,k) ||^2, 0)]
\end{aligned}
\end{equation}
where $\mathbf{1}$ is the indicator function, and $t$ defines a margin that regulates the degree of dissimilarity in semantic asymmetries. 

% \noindent 
\textbf{Overall loss function.} We approach tumor segmentation as a binary segmentation task, utilizing the Dice loss, binary cross-entropy loss, and margin loss as our objective function. The overall loss function is formulated as: $l =  l_{dice} +  l_{ce} + \beta l_{margin},$ where $\beta$ is the weight balancing the different losses.

\subsection{Siamese Segmentation Architecture}
Our SATS architecture comprises the encoder-decoder module and the projection head. While both components are engaged during the training process, only the encoder-decoder module is required during inference.

\textbf{Siamese encoder-decoder.}
The backbone is a shared U-shaped encoder-decoder architecture, as shown in Fig.~\ref{fig:overview}. The encoder employs repeated applications of 3D residual blocks, with each block comprising two convolutional layers with $3\times3\times3$ kernels. Each convolutional layer is succeeded by InstanceNorm normalization and LeakyReLU activation. For downsampling, a convolutional operation with a stride of $2$ is utilized to halve the resolution of the input feature maps. The initial number of filters is $32$ and doubles after each downsampling step to maintain constant time complexity except for the last layer. In total, the encoder performs four downsampling operations.

\textbf{Projection head.} ~We utilize a non-linear projection $g$ to transform the features before calculating the distance in margin loss, which aims to enhance the quality of the learned features. It consists of three $1\times1\times1$ convolution layers with $16$ channels followed by a unit-normalization layer. The first two layers in the projection head use the ReLU activation function. We hypothesize that directly applying metric learning to segmentation features might lead to information loss and diminish the model's effectiveness. For example, some asymmetries in CT images are non-pathological and may stem from variations in the patient's head positioning and posing, yet they are beneficial for segmentation. Utilizing a non-linear projection may filter out such irrelevant information from the metric learning process, ensuring it is preserved in the features used for segmentation.

%% file: sec/3_experiment.tex
\section{Experiments}

\subsection{Data Preparation and Implementation Details}

\begin{figure}[!b]
\begin{center}
\includegraphics[width=0.6\linewidth]{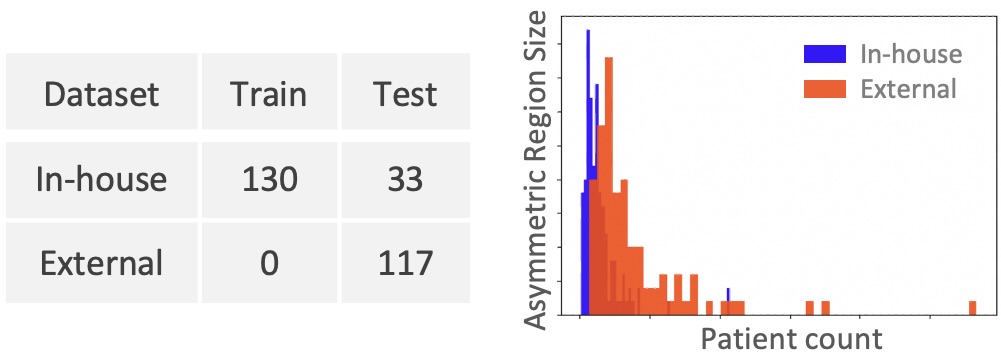}
\end{center}
\caption{Left: the data partitioning situation of the main comparing experiment.
Right: the size of the asymmetric regions of different individuals on the two datasets, with the x-axis displaying each test object.}
\label{Fig:data_analysis}
\end{figure} 

We collected an \emph{in-house dataset} from the hospital for the model development, which consisted of 163 NPC patients with pCT, contrast-enhanced diagnostic CT, and diagnostic MRIs of T1 $\&$ T2 phases. 
Diagnostic CT and MRI were registered as, initially, a rigid transformation~\cite{abs-2307-03535} was applied to the MRI images to approximately align with the CT images. Then, deformable registration algorithm, deeds~\cite{heinrich2012globally}, was utilized to achieve precise alignment. The contrast-enhanced CT and MRIs are used to guide radiation oncologists to generate ground-truth GTV in pCT. 
Also, we collected a public dataset, SegRap2023~\footnote{\url{https://segrap2023.grand-challenge.org/dataset/}}, as~\emph{external testing dataset}, containing 118 non-contrast pCT and enhanced CT. 
\emph{Annotations of all datasets were examined and edited by two experienced radiation oncologists following the international GTV delineation consensus guideline~\cite{lee2018international}}. 
For evaluation, 20 $\%$ of the in-house dataset was randomly selected as the internal testing set, and the entire curated SegRap2023 was used as the external testing dataset.
As shown in Figure~\ref{Fig:data_analysis}, the asymmetric regions in the external data are larger than those of in-house, making the task more challenging.
\textbf{Implementation. }
The model training is divided into two stages. 
In the first stage, only the Siamese encoder-decoder is trained for $800$ epochs with a learning rate of $1e-2$ and decayed via a polynomial schedule.
Then, the projection head is trained for $200$ epochs, with a learning rate of $1e-2$ for the projection head and $1e-5$ for the encoder-decoder, both with decayed via a polynomial schedule.
The patch size is $56\times192\times192$ and the batch size is $2$. 
For the voxel-wise contrastive loss, we use a margin hyperparameter $t=20$ and $\beta=1$.

\subsection{Comparing to State-of-the-art Methods}

\textbf{Comparison methods. }
We conducted a comprehensive comparison of our method with $\textbf{ten}$ cutting-edge approaches, encompassing prominent CNN-based, Transformer-based and Mamba-based methods, to evaluate its performance.
CNN-based methods include STU-Net S~\cite{abs-2304-06716}, STU-Net B~\cite{abs-2304-06716}, STU-Net L~\cite{abs-2304-06716}, MedNeXt~\cite{RoyKUBPIJM23} and nnUNet~\cite{isensee2024nnu}.
Transformer-based methods include UNETR~\cite{HatamizadehTN0M22}, TransUNet~\cite{chen2024transunet}, SwinUNETR~\cite{HatamizadehNTYR21} and its variant SwinUNETR-v2~\cite{HeNYTMX23}.
Mamba-based methods include UMambaBot~\cite{abs-2401-04722}.
To maintain a fair comparison, we trained all competing models for an equal number of epochs, $1000$.
\textbf{Evaluation metrics. }
We evaluate the performance using the Dice similarity coefficient, DSC ($\%$), and the 95th percentile of the Hausdorff distance (HD95, $mm$) and average surface distance (ASD, $mm$) across all cases.

\begin{table*}[!t]
	\centering
    \caption{Quantitative results on NPC GTV segmentation task. In-house$_{train}$ $\Rightarrow$ In-house$_{test}$ represents training on scans from the In-house dataset and segmenting images in the test set of the In-house dataset.  $\uparrow$: Higher values are better. $\downarrow$: Lower values are better. The last column presents the number of model parameters (in millions). The best-performing results are shown in bold while the second-best results are indicated by underlining. $\ddag$: Statistical significant with $P<0.05$ in comparison to our SATS.}
	\resizebox{1\textwidth}{!}{%
		\begin{tabular}{cccccc c}
			\toprule[1.5pt]
			\multirow{2}{*}{Method} & \multicolumn{2}{c}{In-house$_{train}$ $\Rightarrow$ In-house$_{test}$} & \multicolumn{2}{c}{In-house$_{train}$ $\Rightarrow$ External$_{test}$} & \multirow{2}{*}{Para. (M)} \\
			\cmidrule(lr){2-3} \cmidrule(lr){4-5} 
			~ &  DSC $\uparrow$ & HD95 $\downarrow$ &  DSC $\uparrow$ & HD95 $\downarrow$  \\
			\midrule[1pt]
            UMambaBot  & 79.27 $\pm$ 7.77$\ddag$  &  4.66 $\pm$ 3.93 & 63.08 $\pm$ 12.02$\ddag$ & 9.22 $\pm$ 7.52  & 64.76  \\
            \midrule[0.5pt]
            UNETR  & 75.75 $\pm$ 8.92$\ddag$  & 5.41 $\pm$ 4.07 & 62.56 $\pm$ 12.50$\ddag$  & 9.27 $\pm$ 7.46  & 93.01 \\
            TransUNet  & 78.95 $\pm$ 8.28$\ddag$  & 6.42 $\pm$ 12.89 & 62.96 $\pm$ 13.49$\ddag$ & 9.52 $\pm$ 8.16  &  119.37\\
            SwinUNETR  & 80.01 $\pm$ 8.04  & 4.52 $\pm$ 2.77  & 62.90 $\pm$ 11.90$\ddag$  & 9.11 $\pm$ 7.41 &  62.19 \\
            SwinUNETR-V2  & \underline{80.41 $\pm$ 7.80}   &  4.17 $\pm$ 2.40  & 63.81 $\pm$ 12.11$\ddag$ & 8.90 $\pm$ 7.32  & 72.89 \\
            \midrule[0.5pt]
            MedNeXt  & 76.15 $\pm$ 9.83$\ddag$  &  5.09 $\pm$ 3.93 &\underline{64.77 $\pm$ 12.05}$\ddag$  &  9.01 $\pm$ 7.50  &  61.80 \\
            STU-Net S  & 79.04 $\pm$ 7.30  & 4.95  $\pm$ 4.08  &63.50 $\pm$ 11.96$\ddag$ &  9.07 $\pm$ 7.33  &  14.60  \\
            STU-Net B  & 78.86 $\pm$ 7.38  &  4.91 $\pm$ 3.98  & 63.54 $\pm$ 12.05$\ddag$ & 9.14 $\pm$ 7.46 &  58.26 \\
            STU-Net L  & 79.24 $\pm$ 7.23  & 4.64 $\pm$ 3.80 & 63.50 $\pm$ 11.91$\ddag$ & 9.09 $\pm$ 7.25  &  440.30 \\
            nnUNet   & 79.30 $\pm$ 9.77  & \underline{4.07 $\pm$ 2.77} & 64.40 $\pm$ 11.82$\ddag$ & \underline{8.84 $\pm$ 7.40} & 30.70 \\
            \midrule[0.5pt]
            SATS (Ours)  & \textbf{81.22 $\pm$ 8.33 }   & \textbf{4.02 $\pm$ 2.74} &  \textbf{66.80 $\pm$ 12.02} & \textbf{8.51 $\pm$ 7.84} & 30.70\\
		\bottomrule[1.5pt]
		\end{tabular}
	}
	\label{tab:main_result_In}
\end{table*}

\begin{figure*}[!t]
\begin{center}
\includegraphics[width=1\linewidth]
{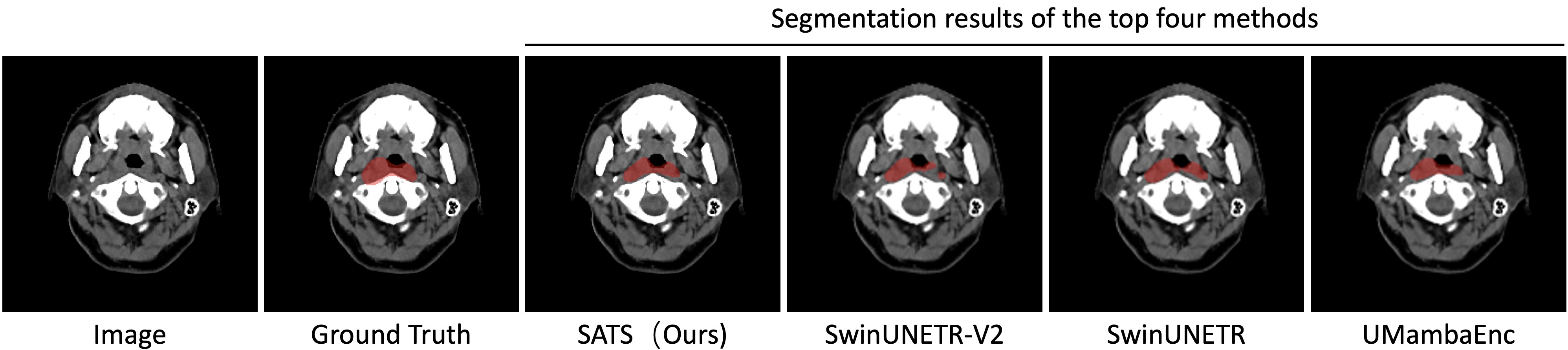} 
\end{center}
\caption{Example CT slices with tumor segmentation overlays (red color) using different methods on the \emph{In-house dataset}. }
\label{Fig_inhouse}
\end{figure*}

\begin{figure}[t]
    \centering
    \begin{minipage}{0.44\textwidth}
        \centering
       \includegraphics[width=\linewidth]{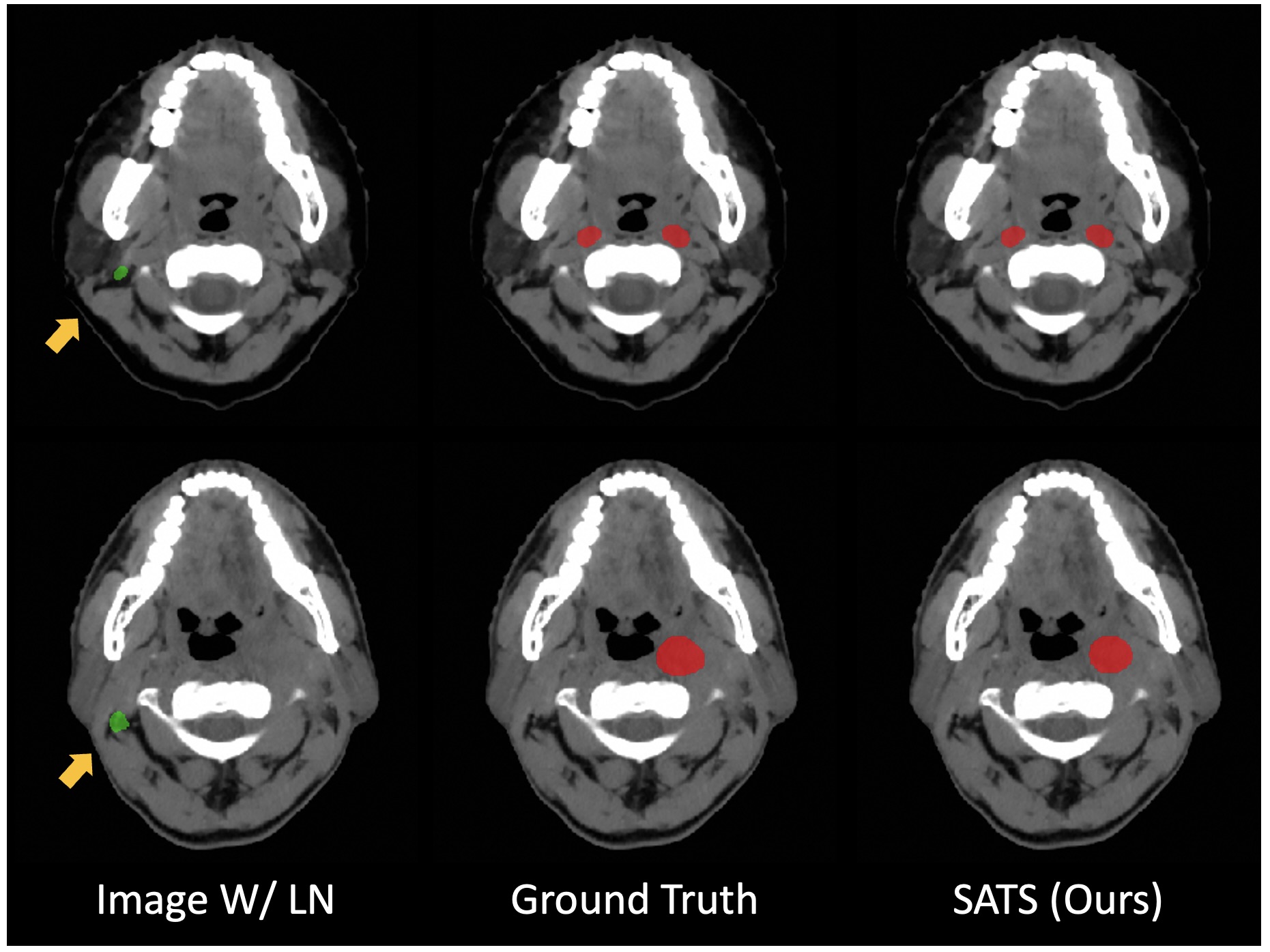} 
        \caption{
        Segmentation results for two nodal-involved (green) patients.}
        \label{fig:robust_LN}
    \end{minipage}\hfill 
    \begin{minipage}{0.54\textwidth}
        \centering
        \includegraphics[width=\linewidth]{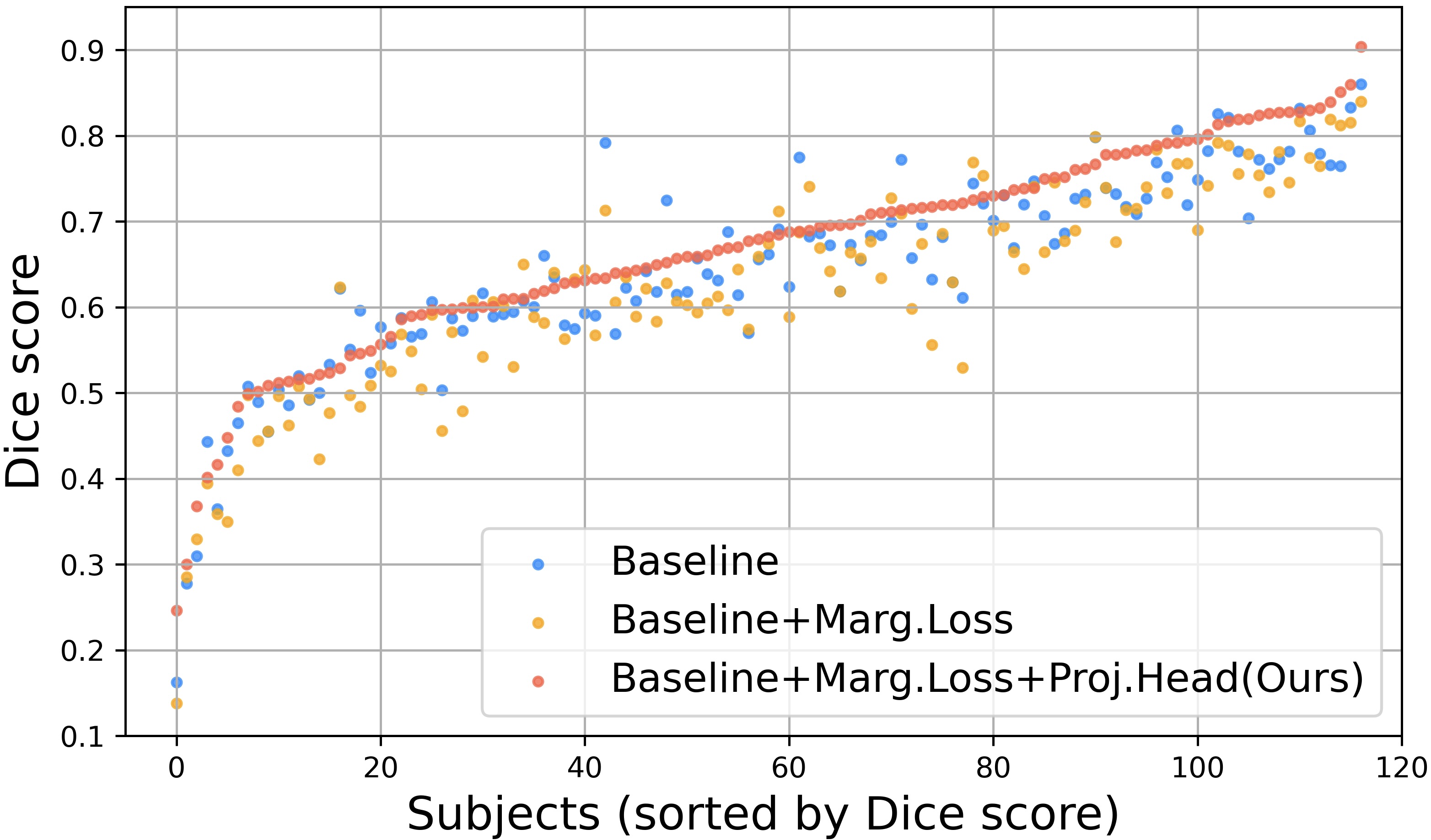}
        \caption{Dice score compared to the baselines shown for each test subject. }
        \label{fig:improvement}
    \end{minipage}
    \label{fig:two_images_AB}
\end{figure}

\textbf{In-house dataset performance.} 
Table.~\ref{tab:main_result_In} summarizes the quantitative segmentation performance and model parameters.  Under a relatively small number of parameters, the proposed SATS demonstrates an improvement over previous approaches. For example, SATS outperforms the transformer-based SWinUNETR-V2 in DSC, and HD95 by 0.81\% and 3.6\%, respectively. Figure~\ref{Fig_inhouse} presents the segmentation results of the top four performing methods (SATS, SwinUNETR-V2, SwinUNETR, and nnUNet) on a sample from the in-house dataset. It can be observed that our SATS method exhibits higher accuracy in boundary segmentation (e.g., the nasal septum). 
\textbf{Robustness.} Large primary tumors can cause asymmetrical anatomical changes. NPC patients often show lymphatic involvement, significantly affecting the integrity and symmetry of nearby structures. Figure~\ref{fig:robust_LN} highlights cases of lymphatic invasion, demonstrating our robustness in handling lymph nodes while accurately segmenting the primary tumor.

\begin{figure*}[!t]
\begin{center}
\includegraphics[width=1\linewidth]{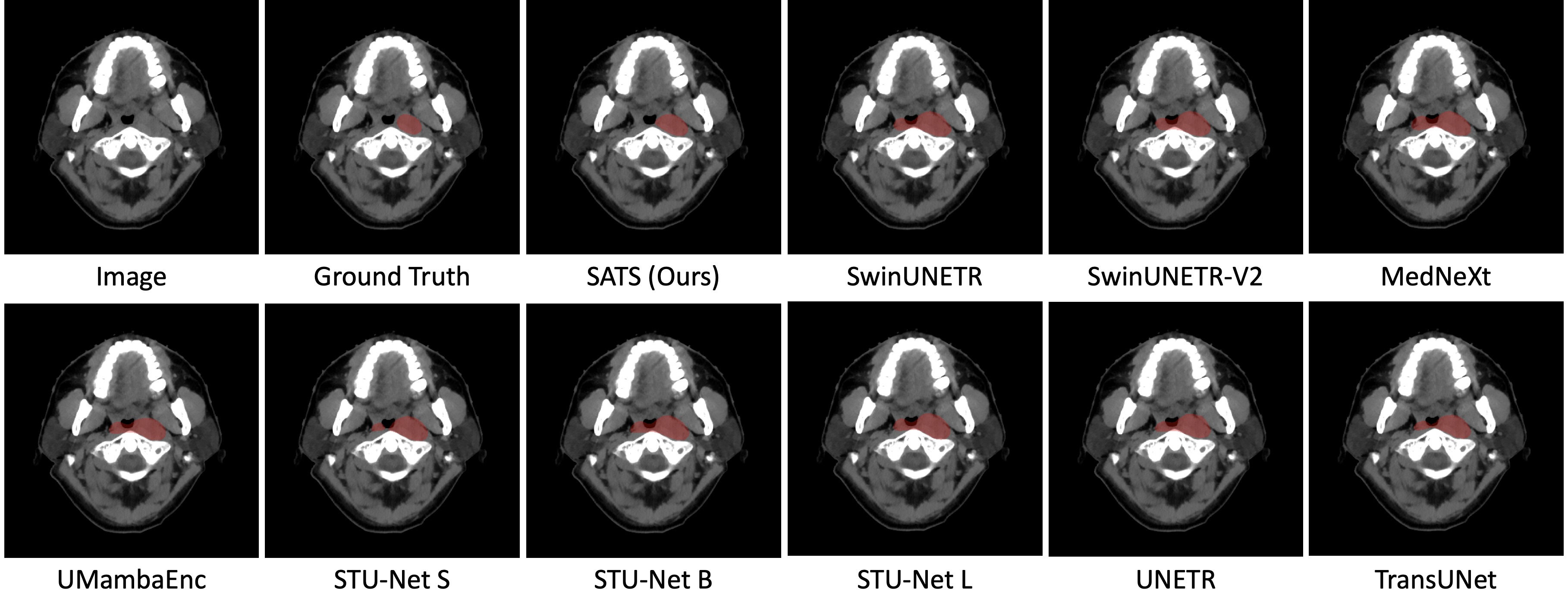}
\end{center}
\caption{Example CT slices with tumor segmentation overlays (red color) using different methods on the \emph{External dataset}. }
\label{Fig_external}
\end{figure*}

\textbf{Performance in external evaluation. } 
Table~\ref{tab:main_result_In} and Figure~\ref{Fig_external} summarize the external testing results. Several conclusions can be drawn. First, the proposed SATS achieves the best performance as compared to all other leading methods in external evaluation. As compared to an increase of 0.92\% DSC over the 2nd best-performing method (nnUNet) in internal testing, SATS exhibits a substantial improvement of 4.4\% DSC over nnUNet in external evaluation. This demonstrates the better generalizability of the proposed semantic asymmetry learning in NPC GTV segmentation.  Third, the proposed SATS consistently outperforms other leading methods in terms of HD95 (>3.7\% error reduction).  Lastly, although SwinUNETR-V2 performs 2nd best in internal testing with 1.11\% DSC improvement over nnUNet, nnUnet outperforms SwinUNETR-V2 in external testing by 0.61\% DSC. This result indicates the strong performance of CNN-based nnUNet over transformer-based segmentation models.

\begin{table}[ht]
\caption{Influence of the effect of the projection head and margin loss. }
	\begin{center}
		\scalebox{1.}{
			\resizebox{0.8\textwidth}{!}{
					\begin{tabular}{cc|ccc}
						\toprule
			Proj. Head & Marg. Loss   & DSC ($\%$) & ASD ($mm$) & HD95 ($mm$) \\ 
						\hline
				$\usym{2717}$ &   $\usym{2717}$   & 63.44 $\pm$ 10.54  &  2.97 $\pm$ 1.37  &  7.22 $\pm$ 3.34     \\ 
                  $\usym{2717}$   & $\checkmark$     & 61.50 $\pm$ 10.02 & 3.20 $\pm$ 1.39  & 7.73  $\pm$ 3.58     \\
                     $\checkmark$ & $\checkmark$    & 66.32 $\pm$ 10.48 &  2.60 $\pm$ 1.36  &  6.58 $\pm$ 3.50     \\
						\bottomrule
						\end{tabular}
					}}
	\end{center}
	\label{tab:ablation}
\end{table}

\subsection{Ablation Studies}
\textbf{Effect of projection head and margin loss. } Table~\ref{tab:ablation} demonstrates performance metrics for different segmentation models on the external data (In-house$_{train}$ $\Rightarrow$ External$_{test}$).
Although the margin loss is append, the model has lower DSC and higher ASD/HD95 than that of the initial model, suggesting that margin loss alone is insufficient to improve performance. 
In contrast, there is a significant performance boost ($+$4.98$\%$ DSC, $-$0.60$mm$ ASD and $-$1.15$mm$ HD95) when the projection head module and margin loss are added. 
\textbf{Effect of semantic asymmetry learning. } In Figure~\ref{fig:improvement}, we present a comparative analysis of our proposed method against the baseline configurations that exclude the projection head module and/or employ margin loss baselines. As depicted, our method achieves the highest Dice score, demonstrating consistent superiority over all baseline models across the majority of the 117 test scans. 
\textbf{Failure cases analysis.}~Our method performs poorly on extreme outliers in Figure~\ref{fig:improvement}, misclassifying symmetric lesions as asymmetric and achieving lower DSC than the nnUNet baseline. A complementary framework combining our approach with nnUNet could enhance clinical robustness.

%% file: Paper-1607.bbl
\begin{thebibliography}{10}
\providecommand{\url}[1]{\texttt{#1}}
\providecommand{\urlprefix}{URL }
\providecommand{\doi}[1]{https://doi.org/#1}

\bibitem{abs-2307-03535}
Bai, X., Bai, F., Huo, X., et~al.: Matching in the wild: Learning anatomical embeddings for multi-modality images. CoRR  \textbf{abs/2307.03535} (2023)

\bibitem{bai2021deep}
Bai, X., Hu, Y., Gong, G., Yin, Y., Xia, Y.: A deep learning approach to segmentation of nasopharyngeal carcinoma using computed tomography. Biomedical Signal Processing and Control  \textbf{64},  102246 (2021)

\bibitem{ChenWZLCHXHLLM20}
Chen, H., Wang, Y., Zheng, K., Li, W., Chang, C., Harrison, A.P., et~al.: Anatomy-aware siamese network: Exploiting semantic asymmetry for accurate pelvic fracture detection in x-ray images. In: ECCV. vol. 12368, pp. 239--255 (2020)

\bibitem{ChenQYLLLGW20}
Chen, H., Qi, Y., Yin, Y., et~al.: Mmfnet: {A} multi-modality {MRI} fusion network for segmentation of nasopharyngeal carcinoma. Neurocomputing  \textbf{394},  27--40 (2020)

\bibitem{chen2024transunet}
Chen, J., Mei, J., Li, X., Lu, Y., Yu, Q., Wei, Q., et~al.: Transunet: Rethinking the u-net architecture design for medical image segmentation through the lens of transformers. Medical Image Analysis p. 103280 (2024)

\bibitem{chen2019nasopharyngeal}
Chen, Y.P., Chan, A.T., Le, Q.T., Blanchard, P., Sun, Y., Ma, J.: Nasopharyngeal carcinoma. The Lancet  \textbf{394}(10192),  64--80 (2019)

\bibitem{chua2016nasopharyngeal}
Chua, M.L., Wee, J.T., Hui, E.P., Chan, A.T.: Nasopharyngeal carcinoma. The Lancet  \textbf{387}(10022),  1012--1024 (2016)

\bibitem{HatamizadehNTYR21}
Hatamizadeh, A., Nath, V., Tang, Y., Yang, D., Roth, H.R., Xu, D.: Swin {UNETR:} swin transformers for semantic segmentation of brain tumors in {MRI} images. In: Brainlesion: Glioma, Multiple Sclerosis, Stroke and Traumatic Brain Injuries. vol. 12962, pp. 272--284 (2021)

\bibitem{HatamizadehTN0M22}
Hatamizadeh, A., Tang, Y., Nath, V., Yang, D., Myronenko, A., Landman, B.A., et~al.: {UNETR:} transformers for 3d medical image segmentation. In: {IEEE} Winter Conference on Applications of Computer Vision. pp. 1748--1758 (2022)

\bibitem{HeNYTMX23}
He, Y., Nath, V., Yang, D., Tang, Y., Myronenko, A., Xu, D.: Swinunetr-v2: Stronger swin transformers with stagewise convolutions for 3d medical image segmentation. In: MICCAI. vol. 14223, pp. 416--426 (2023)

\bibitem{heinrich2012globally}
Heinrich, M.P., Jenkinson, M., Brady, S.M., Schnabel, J.A.: Globally optimal deformable registration on a minimum spanning tree using dense displacement sampling. In: MICCAI. pp. 115--122 (2012)

\bibitem{huang2019achieving}
Huang, J.b., Zhuo, E., Li, H., Liu, L., Cai, H., Ou, Y.: Achieving accurate segmentation of nasopharyngeal carcinoma in mr images through recurrent attention. In: MICCAI. pp. 494--502 (2019)

\bibitem{HuangLLW22}
Huang, J., Li, H., Li, G., Wan, X.: Attentive symmetric autoencoder for brain {MRI} segmentation. In: MICCAI. vol. 13435, pp. 203--213 (2022)

\bibitem{huang2024lidia}
Huang, W., Liu, W., Zhang, X., Yin, X., Han, X., Li, C., Gao, Y., et~al.: Lidia: Precise liver tumor diagnosis on multi-phase contrast-enhanced ct via iterative fusion and asymmetric contrastive learning. In: MICCAI. pp. 394--404 (2024)

\bibitem{abs-2304-06716}
Huang, Z., Wang, H., Deng, Z., Ye, J., Su, Y., et~al.: Stu-net: Scalable and transferable medical image segmentation models empowered by large-scale supervised pre-training. CoRR  \textbf{abs/2304.06716} (2023)

\bibitem{isensee2024nnu}
Isensee, F., Wald, T., Ulrich, C., Baumgartner, M., Roy, S., Maier-Hein, K., Jaeger, P.F.: nnu-net revisited: A call for rigorous validation in 3d medical image segmentation. In: MICCAI. pp. 488--498 (2024)

\bibitem{ke2020development}
Ke, L., Deng, Y., Xia, W., Qiang, M., et~al.: Development of a self-constrained 3d densenet model in automatic detection and segmentation of nasopharyngeal carcinoma using magnetic resonance images. Oral Oncology  \textbf{110},  104862 (2020)

\bibitem{lee2018international}
Lee, A.W., Ng, W.T., Pan, J.J., Poh, S.S., Ahn, Y.C., AlHussain, H.o.: International guideline for the delineation of the clinical target volumes (ctv) for nasopharyngeal carcinoma. Radiotherapy and Oncology  \textbf{126}(1),  25--36 (2018)

\bibitem{li2024improved}
Li, C., Zhang, X., Gao, Y., Yin, X., Lu, L., et~al.: Improved esophageal varices assessment from non-contrast ct scans. In: MICCAI. pp. 349--359 (2024)

\bibitem{li2019tumor}
Li, S., Xiao, J., He, L., Peng, X., Yuan, X.: The tumor target segmentation of nasopharyngeal cancer in ct images based on deep learning methods. Technology in cancer research \& treatment  \textbf{18},  153--160 (2019)

\bibitem{9684475}
Li, Y., Dan, T., Li, H., Chen, J., Peng, H., Liu, L., Cai, H.: Npcnet: Jointly segment primary nasopharyngeal carcinoma tumors and metastatic lymph nodes in mr images. IEEE Transactions on Medical Imaging  \textbf{41}(7),  1639--1650 (2022)

\bibitem{liao2022automatic}
Liao, W., He, J., Luo, X., Wu, M., Shen, Y., et~al.: Automatic delineation of gross tumor volume based on magnetic resonance imaging by performing a novel semisupervised learning framework in nasopharyngeal carcinoma. International Journal of Radiation Oncology* Biology* Physics  \textbf{113}(4),  893--902 (2022)

\bibitem{liu2019using}
Liu, C.F., Padhy, S., Ramachandran, S., et~al.: Using deep siamese neural networks for detection of brain asymmetries associated with alzheimer's disease and mild cognitive impairment. Magnetic resonance imaging  \textbf{64},  190--199 (2019)

\bibitem{LiuZZLZZLWY19}
Liu, Y., Zhou, Z., Zhang, S., Luo, L., Zhang, Q., Zhang, F., et~al.: From unilateral to bilateral learning: Detecting mammogram masses with contrasted bilateral network. In: MICCAI. vol. 11769, pp. 477--485 (2019)

\bibitem{luo2021efficient}
Luo, X., Liao, W., Chen, J., Song, T., Chen, Y., Zhang, S., et~al.: Efficient semi-supervised gross target volume of nasopharyngeal carcinoma segmentation via uncertainty rectified pyramid consistency. In: MICCAI. pp. 318--329 (2021)

\bibitem{luo2023deep}
Luo, X., Liao, W., He, Y., Tang, F., Wu, M., Shen, Y., et~al.: Deep learning-based accurate delineation of primary gross tumor volume of nasopharyngeal carcinoma on heterogeneous magnetic resonance imaging: a large-scale and multi-center study. Radiotherapy and Oncology p. 109480 (2023)

\bibitem{abs-2401-04722}
Ma, J., Li, F., Wang, B.: U-mamba: Enhancing long-range dependency for biomedical image segmentation. CoRR  \textbf{abs/2401.04722} (2024)

\bibitem{ma2019nasopharyngeal}
Ma, Z., Zhou, S., Wu, X., Zhang, H., Yan, W., et~al.: Nasopharyngeal carcinoma segmentation based on enhanced convolutional neural networks using multi-modal metric learning. Physics in Medicine \& Biology  \textbf{64}(2),  025005 (2019)

\bibitem{mei2021automatic}
Mei, H., Lei, W., Gu, R., Ye, S., Sun, Z., Zhang, S., et~al.: Automatic segmentation of gross target volume of nasopharynx cancer using ensemble of multiscale deep neural networks with spatial attention. Neurocomputing  \textbf{438},  211--222 (2021)

\bibitem{men2017deep}
Men, K., Chen, X., Zhang, Y., Zhang, T., Dai, J., Yi, J., Li, Y.: Deep deconvolutional neural network for target segmentation of nasopharyngeal cancer in planning computed tomography images. Frontiers in oncology  \textbf{7}, ~315 (2017)

\bibitem{razek2012mri}
Razek, A.A.K.A., King, A.: Mri and ct of nasopharyngeal carcinoma. American Journal of Roentgenology  \textbf{198}(1),  11--18 (2012)

\bibitem{RoyKUBPIJM23}
Roy, S., K{\"{o}}hler, G., Ulrich, C., Baumgartner, M., Petersen, J., Isensee, F., et~al.: Mednext: Transformer-driven scaling of convnets for medical image segmentation. In: MICCAI. vol. 14223, pp. 405--415 (2023)

\bibitem{Tian2023SAMEAS}
Tian, L., Li, Z., Liu, F., Bai, X., Ge, J., Lu, L., et~al.: Same++: A self-supervised anatomical embeddings enhanced medical image registration framework using stable sampling and regularized transformation. ArXiv  \textbf{abs/2311.14986} (2023)

\bibitem{zeng2023two}
Zeng, B., Wang, H., Xu, J., Tu, P., Joskowicz, L., Chen, X.: Two-stage structure-focused contrastive learning for automatic identification and localization of complex pelvic fractures. IEEE Transactions on Medical Imaging  \textbf{42}(9),  2751--2762 (2023)

\end{thebibliography}
